\documentclass[a4paper]{PoS}
\usepackage{siunitx}
\usepackage{eurosym}
\usepackage[switch, modulo]{lineno}

\title{Design study of an air-Cherenkov telescope for harsh environments with efficient air-shower detection at 100 TeV.}

\ShortTitle{ACT for the South Pole}

\author{\speaker{Jan Auffenberg} $^{1}$, Thomas Bretz $^{1}$, Lukas Middendorf $^{1}$, Tim Niggemann $^{1}$, Leif R\"adel $^{1}$,  Merlin Schaufel $^{1}$, Sebastian Schoenen $^{1}$, Johannes Schumacher $^{1}$, Christopher Wiebusch $^{1}$
\\
{\it 
$^1$ RWTH Aachen University\\
}
E-mail: \email{jauffenb@icecube.wisc.edu}
}

\abstract{Telescopes, designed with semi-conductor based photo sensors, have the potential to detect Cherenkov or fluorescence light emitted by cosmic-rays in the atmosphere. Such telescopes promise a high duty cycle and efficiency in remote harsh environments. Given the relatively low costs and robustness of these instruments, this technology could prove interesting especially if deployed in large numbers with existing and future extended cosmic-ray and gamma ray detectors, including the Pierre Auger observatory, HAWC, IceCube and CTA. They may have the potential to enhance the sensitivity of these instruments for the detection of high-energy gamma rays and cosmic-ray air showers. In addition, for neutrino telescopes such a technology could prove to be an efficient cosmic-ray veto on the surface. In this contribution the current motivation, design, and development of a prototype SiPM based air Cherenkov telescope is described. The results of initial sensitivity studies, and the readiness of the system for first tests, including those proposed for the South Pole are shown.}

\FullConference{The 34th International Cosmic Ray Conference,\\
		30 July- 6 August, 2015\\
		The Hague, The Netherlands}

\begin{document}

\section{Introduction}

Recently published results show that IceCube \cite{IceCube} has measured an astrophysical neutrino flux with very high significance but no evidence for a point source has been found \cite{4YHese} so far. In the region above $\SI{100} {\tera \electronvolt}$ primary neutrino energy the measured diffuse flux is significantly above the background of cosmic-ray induced neutrinos and muons. One of the main tasks of an extended IceCube detector, \textit{IceCube-Gen2}, will be the detection and high-quality reconstruction of a sufficient number of astrophysical neutrino events for astronomical observations and the measurement of the corresponding neutrino spectrum with high precision (see also \cite{SurV, Gen2}).\\
The main backgrounds for extraterrestrial neutrino detection are cosmic-ray induced particles. One way to suppress these backgrounds in IceCube is to look for particles that traversed the whole Earth, leaving only neutrinos - astrophysical and atmospheric. In the southern sky, the cosmic-ray induced muon background is dominant because high-energy muons are not absorbed within at least 2\,km overburden of ice above the detector. However, cosmic-ray induced air showers can be detected with dedicated air shower arrays at the surface \cite{IceTop}. Such a surface detector can be used to veto the cosmic-ray muons reaching the deep in-ice detector. To some extend, such a veto detector is able to select and suppress even the atmospheric neutrino background as they are also accompanied by air showers on the surface. In fact, IceTop is already used today in several analyses to suppress the atmospheric background and increase the sensitivity of IceCube for astrophysical neutrinos in the southern sky \cite{EHE,PS}. The properties of IceTop as a veto motivate further studies of a specialized detector to very efficiently detect cosmic-ray muons already at the surface. This could make it possible to select neutrinos that interacted anywhere in the entire ice sheet above the IceCube in-ice detector as astrophysical neutrino candidates. The possible impact of a surface veto detector to astrophysical neutrino measurements with IceCube is discussed in \cite{SurV}.\\
Detector extensions to veto cosmic-ray induced signals on the surface have to be very efficient in different ways:
\begin{itemize}
\item A low energy threshold and high detection efficiency of a surface veto is strongly related to the detectable neutrino flux.
\item A large duty cycle is needed for a surface veto detector to detect as many astrophysical events as possible.
\item The number of detectable astrophysical events increases linear with the azimuthal coverage of a surface array. With increasing declination the length of the active Volume (ice) increases. Thus a neutrino will more likely interact with increasing declination and produce a signal in the in-ice detector. In the vertical case, only ice directly above the detector (about a factor of 2.5) is gained, while e.g. in the direction of the Galactic Center (declination $\theta = \ang{68}$) the active volume is already about six times larger.
\item The detection system has to be easy to deploy and operate. 
\end{itemize}

The most obvious option for a surface veto are particle detectors which measure the Cherenkov- or scintillation light produced in an enclosed active volume. First simulations based on shower parameterizations underline the importance of large detection volumes and sensitivity to the electromagnetic and the muonic component of the air shower \cite{SurfaceExtSim}. A large uncertainty are intrinsic air shower fluctuations and will need careful investigation in the future.

In the following, an air Cherenkov telescope is discussed as an alternative detection method which uses the atmosphere as active volume. First, the estimated distribution of air Cherenkov telescopes and its technical properties is discussed that would be needed for a surface veto detector. After that, the prototype telescope IceAct is described that has been build for the South Pole. In addition, we show measurements from first field tests with IceAct and summarize.

\section{Properties of an air Cherenkov telescope array to efficiently detect and veto air showers at the South Pole.}
\label{IceVeto}

Efficiently running an air Cherenkov telescope at the South Pole requires special demands on the properties for a single detector. In order to use such telescopes to veto air showers constrain the needed properties further. In the following we will briefly discuss the required properties of a single detector and a whole detector array that could be operated at the South Pole as a surface veto for cosmic-ray air showers.

\subsection{Constrains to veto detectors at the surface on the energy threshold and the detection efficiency}

A surface detector on top of IceCube to veto air shower induced signals in the in-ice detector could improve the astrophysical neutrino search of IceCube in various ways. Especially the detection of well reconstructed high energy neutrinos could be improved with a surface veto detector \cite{SurV}.

For instance, a veto detector that would efficiently reduce the air shower background at $\SI{10}{\tera\electronvolt}$ muon energy in the deep in-ice IceCube detector covering the entire southern sky (\ang{0}-\ang{85} declination) would be able to detect about 40 astrophysical neutrinos per year \cite{SurV}. This would double the number of astrophysical neutrinos that IceCube can detect in this energy region. The energy of an astrophysical neutrino is smaller than the primary energy of an air shower that produces the same signature in the in-ice detector.
As a result, the detection efficiency of cosmic-ray air showers for a detector should be very efficient at $> \SI{100}{\tera\electronvolt}$ depending on, e.g, inclination or particle ID. The detection of cosmic rays with $\SI{100}{\tera\electronvolt}$ primary energy is set as benchmark for the prototype telescope IceAct.

\subsection{Environmental constrains for an air Cherenkov telescope at the South Pole}

The environmental conditions at the geographic South Pole are harsh. Temperatures between $\SI{-13}{\celsius}$ and $\SI{-82}{\celsius}$ are measured and wind speeds of up to $\SI{25}{\meter/\second}$ prevail. In addition, one can observe a continuous drift of micro crystalline snow on the surface. Together with an almost completely dry atmosphere the overall conditions are very hard to simulate in laboratories but have to be considered when developing a telescope.
Other environmental parameters like the darkness and transparency of the sky are also essential parameters for the efficient operation of a Cherenkov telescope. 
Analyses of LIDAR measurements at the Amundsen Scott South Pole station suggest the overall time of sufficient weather conditions for measurements with an air Cherenkov telescope to be on the order of $20\%-25\%$ of the year \cite{Segev}. Tests with a prototype at the South Pole will have to show how much impact moon or aurora australis have on the observation time.

\subsection{Geometrical constrains for an air Cherenkov array}

Based on current knowledge we estimate the cost for two different detector configurations. One is covering the same space as IceTop used as a veto, the other is covering the space to veto air-showers from the direction of the Galactic Center. 
\begin{figure}[htbp]
\begin{center}
\includegraphics[width=2.9in]{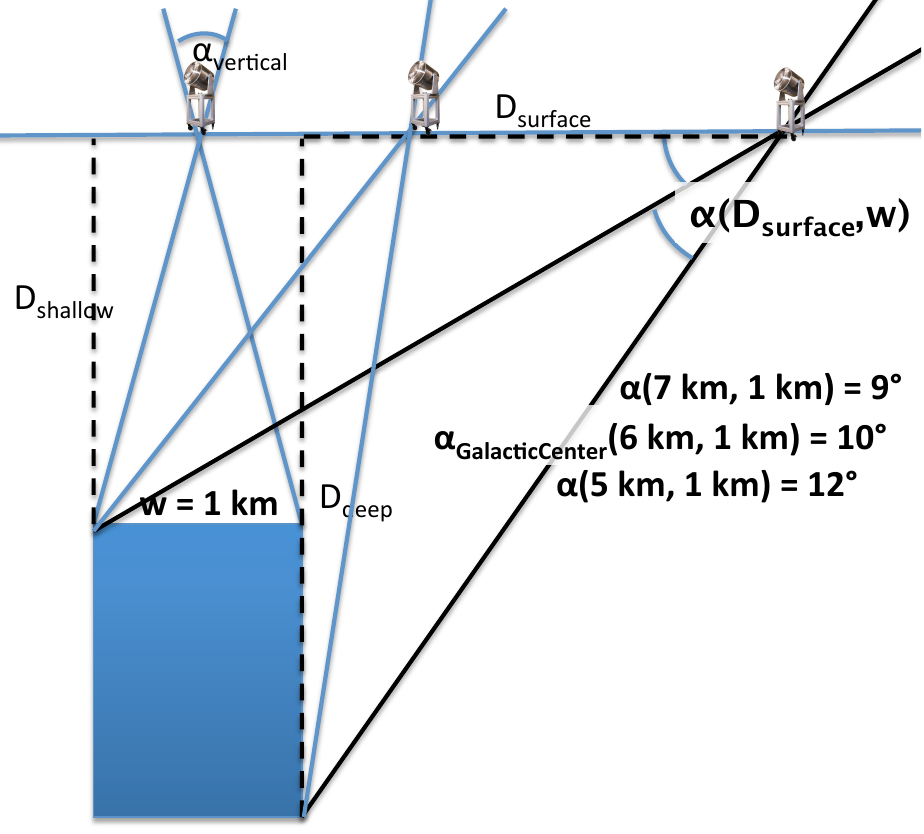}
\caption{Discussion of the geometry of a surface detector to veto cosmic-rays, based on air Cherenkov telescopes. The blue rectangle symbolizes the  IceCube in-ice detector. Assuming a cylindrical shape of the IceCube detector the demanded field of view on the air Cerenkov telescope detectors for different declination can be geometrically calculated. Where the necessary field of view for the vertical case is as large as $\ang{37}$ in $\SI{7}{\kilo\meter}$ distance a field of view of $\ang{9}$ will be sufficient to cover the needed amount of the sky.}
\label{IceVetoGeo}
\end{center}
\end{figure} 
 
Due to design, air Cherenkov telescopes can only observe a small fraction of the night sky at once. However, for a veto detector, where the showers of interest have to traverse to the in-ice detector, only a limited region of the sky needs to be observed by a single telescope. Figure~\ref{IceVetoGeo} illustrates the underlying geometrical properties faced by an extended surface veto array. The blue rectangle is the approximately cylindrical shape of the IceCube in-ice detector. The depth of the in-ice detector $D_{\mathrm{shallow}}$ is $\SI{1.5}{\kilo\meter}$ and the diameter $W$ is chosen as $\SI{1}{\kilo\meter}$ (about the size of IceCube).
We want cover the space of all possible air showers that pass IceTop and IceCube. To cover the central vertical overburden of IceCube, a single telescope array  with a field of view of $\alpha_{\mathrm{vertical}} = \ang{37}$ is needed (see Figure \ref{IceVetoGeo}).
As this central telescope array can only detect air showers with more than $E_{\mathrm{prim}}>\SI{100}{TeV}$ primary energy in less than $300 \si{\meter}$ distance, we need a ring of 7 additional telescope arrays in about $200 \si{\meter}$ distance. In sum we end up with 8 telescope arrays to cover the same space like IceTop as a veto for IceCube.
With the air Cherenkov telescope prototype IceAct that has a field of view of $\ang{12}$ in each of the arrays seven telescopes are needed. With a field of view of a single pixel to be $\Theta_{\mathrm{pixel}}=\ang{1.5}$, this results in 430 channels that are needed in each array. Thus, in sum, about 3440 channels are needed to cover every air shower direction that is contained in IceTop and IceCube. Based on table \ref{cost}, which summarizes the main items in cost of the hardware of the prototype telescope, the cost per channel can be estimated to $85$ \euro{}.
Accounting for contingency and items not stated, we calculate with $170$ \euro{} per channel, the total investment will be about $650\,000$ \euro{} for an air Cherenkov detector that covers IceTop. Both, the cost per channel and the total number of channels are only roughly estimated here.
\begin{table}[ht!]
\centering
\begin{tabular}{||c||c | c | c | c ||}
\hline
 Item & Number & price for one & total price & comment\\
\hline \hline \hline
\textbf{SIPMs} & & & \textbf{ 1472\euro{}} &\\
 $6\times6\,\si{mm^2} $, $50\,\mu m$ & 64 &  23 \euro{} & 1472 \euro{}& SensL C series  \\
\hline
\textbf{Mechanics} &&& \textbf{ 2005 \euro{}} &\\
Fresnel lens & 1 & 100 \euro{} & 100 \euro{}& \\
Glass plane & 1 &  30 \euro{} & 30 \euro{}& \\
Lens tube & 1 &  200 \euro{}& 200 \euro{}& \\
Stand & 1 & 100 \euro{} & 100 \euro{}& \\
Focal plane & 1 & 50 \euro{} & 50 \euro{}& \\
Winston cones & 61 & 25 \euro{} & 1525 \euro{}& \\
Filter & 61 & 1 \euro{} & 61 \euro{} & \\
\hline \hline
\textbf{Electronic}& & & \textbf{3250 \euro{}}&\\
Base board & 1 & 50 \euro{}& 50 \euro{}& \\
Power supply & 64 & 10 \euro{}& 640 \euro{}&\\
Data acquisition & 64 & 40 \euro{}& 2560 \euro{}& based on TARGET 7\\
\hline \hline
Others & & & \textbf{ 18 \euro{}} &  \\
Koax & 0 & 32 \euro{} & 0 \euro{}& price per m \\
Network cable & 2 & 4 \euro{}& 8 \euro{}&\\
ICE Power socket & 1 & 5 \euro{} & 5 \euro{}& \\
Network cable socket & 1 & 5 \euro{}& 5 \euro{}& \\
\hline \hline
\end{tabular}
\caption{Estimated cost of a single IceAct prototype telescope when buying 50. This can get translated into a cost of about $ 85 \euro{}$/channel.}
\label{cost}
\end{table}

For an extended array many more channels are needed as one has to position a telescope every $\SI{200}{\meter}$ on the ring around the in-ice detector. For example, to completely cover the $\ang{2}$ region around the Galactic Center, which is $\ang{28}$ above the horizon, a ring of telescopes in about $\SI{6}{\kilo\meter}$ distance with a field of view of $\ang{12}$ is required. This corresponds to 2420 channels or 450\,000 \euro{}. For all these estimates, it is easy to imagine that the final geometry needs an even larger number of pixels for contingency. In the case of an outer ring, the cabling for power and data transmission cannot be neglected anymore and would most likely more than double the price of the telescopes. The example of one ring around the Galactic Center can be interpreted as an infill array. After the detection of an astrophysical point source in one direction, a dense air Cherenkov telescope array could help to measure the neutrino spectrum of the neutrino source. If adding more than one ring, one should keep in mind that the distance between two rings does not need to be $\SI{200}{meter}$. Due to the projection of the light emission cone $c_s$ on the surface $p_p$ the distance increases with $p_p = c_s/\tan(\phi)$ where $\phi$ is the zenith angle of the incoming air shower that needs to get detected.
These rough estimates show that the price of an air Cherenkov telescope array of this size is not completely unrealistic but needs to be carefully studied. 

\section{A prototype air Cherenkov telescope to run at the South Pole: IceAct}

Based on the constrains described in section \ref{IceVeto}, a prototype was developed which is suited for running at the South Pole. The resulting telescope is based on the design of the FAMOUS telescope, a prototype to estimate the performance of an SiPM based camera for usage in the fluorescence detector of the Pierre Auger Observatory \cite{Famous}. It has been influenced by the proof of concept from FACT \cite{FACT} that SiPMs can be used in air Cherenkov telescopes.
A special power supply was developed that fulfills South Pole ratings \cite{PowerSupply}.\\
The total cost of the telescope is estimated to be far less than 10\,000 \euro{} as summarized in Table \ref{cost}. Here, the largest costs for a single IceAct prototype are summarized. An estimate for cost reduction when building 50 telescopes is included. Infrastructural costs like the cabling for power supply and data transfer or the data storage are not included.

\subsection{The mechanical part of the IceAct prototype}

\begin{figure}[htbp]
\begin{center}
\begin{minipage}[t]{.52\textwidth}
  \centering
  \includegraphics[width=.42\linewidth]{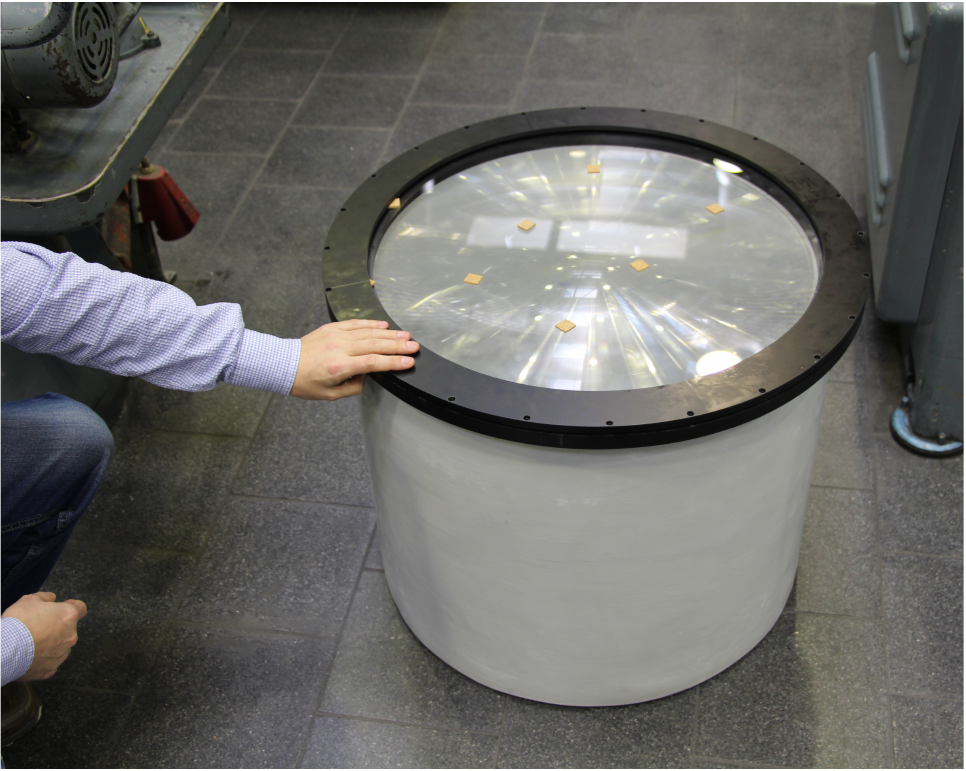}
\caption{Picture of an IceAct prototype telescope that could get deployed at the South Pole in the future. Together with the wooden stand the telescope can stand e.g. on the roof of a building. The telescope has pixels for an \ang{4} field of view and can look from \ang{45} above zenith to vertical.}
  \label{bild_teleskop}
\end{minipage}%
\hfill
\begin{minipage}[t]{.4\textwidth}
  \centering
  \includegraphics[width=.6\linewidth] {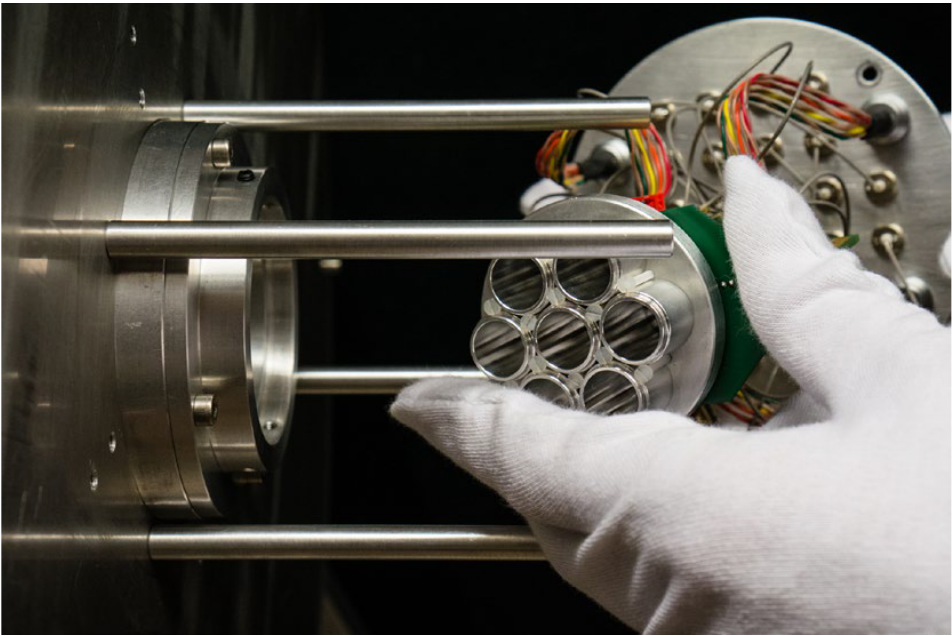}
  \caption{Picture of the 7 pixel camera that will be used in the IceAct prototype. Behind the 7 Winston cones theres are the low noise amplifiers. To connect the power supply and the signal output, different LEMO connectors are used.  }
  \label{Camera_picture}
\end{minipage}
\hfill
\end{center}
\end{figure}

Figure \ref{bild_teleskop} is a picture of the air Cherenkov prototype telescope that has been build to withstand South Pole conditions. Following the optical path of an incoming photon the telescope has the following optical components.

First, it passes a $97\%$ UV-transparent glass plate that protects the entire following system from small snow crystals. It is followed by a Fresnel lens with a $\ang{12}$ field of view and a diameter of $\SI{507}{\mm}$. The carbon cylinder keeps the distance of the lens to the SiPM camera (7 pixels in our prototype) to $\SI{514}{\mm}$. In this distance, seven round Winston cones (c.f. figure \ref{Camera_picture}) guide the light onto the quadratic SiPMs which are each covered by a filter glass plate made of Schott UG11 similar to the filter glass used in the fluorescence detectors of the Pierre Auger Observatory or the FAMOUS prototype \cite{Famous}. 

\subsection{The electronic part of the IceAct prototype}

The electronics of the IceAct prototype are also discussed from the perspective of an incoming photon onto an SiPM. The electrical pulse caused by a single or a bundle of photons is amplified by a $470\, \Omega$ transimpedance amplifier and afterwards is several $\si{\mV}$ in size. DRS\,4 evaluation boards are used to digitize the electrical pulses. The central pixel is read out by both of the two evaluation boards, every other of the six remaining pixels gets one channel each. A two channel coincidence is built within a dynamic time window of up to $\SI{500}{ns}$ in $\SI{10}{\ns}$ steps. Events of all remaining pixels are read out with a speed of up to $\SI{250}{\Hz}$ through USB\,2.0 onto a mini PC. The power supply for the SiPMs has to power each pixel with a different voltage based on the temperature of the SiPM. This dynamic system is described in more detail in \cite{PowerSupply}.

\section{First field test of the prototype telescope}

\begin{figure}[htbp]
\begin{center}
\begin{minipage}[t]{.5\textwidth}
  \centering
  \includegraphics[width=.9\linewidth]{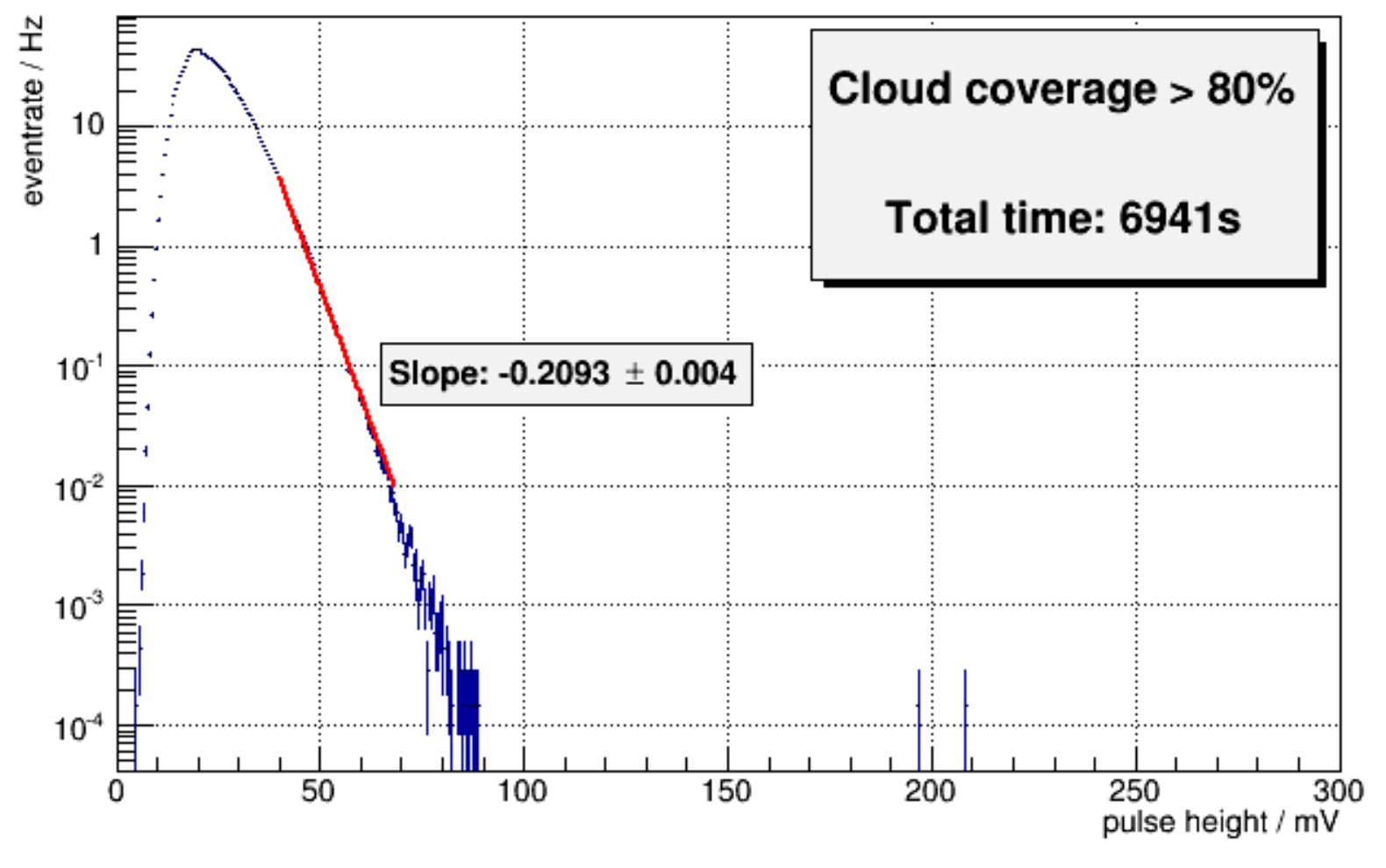}
\caption{Measurement of 2 pixel coincidences with a prototype IceAct telescope near Aachen, Germany. The event rate against the pixel with the largest pulse in the telescope is shown for a measurement with $80 \%$ cloud coverage. No events with a signature expected from air showers like shown in Figure \protect\ref{Event1} were found.}
  \label{cloudy}
\end{minipage}%
\hfill
\begin{minipage}[t]{.47\textwidth}
  \centering
  \includegraphics[width=.9\linewidth] {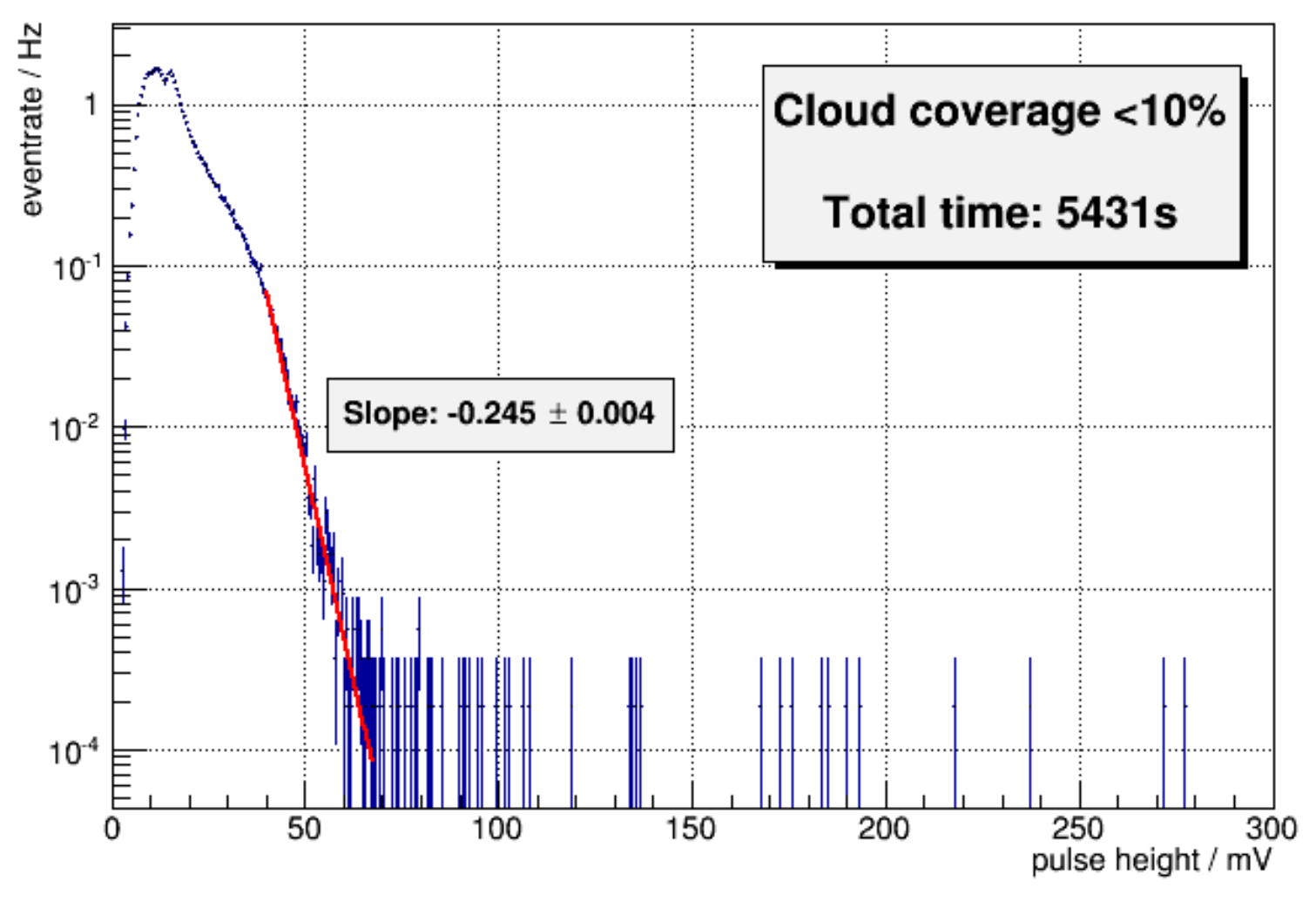}
  \caption{Measurement of 2 pixel coincidences with a prototype IceAct telescope near Aachen. The event rate against the the pixel with the largest pulse in the telescope is shown for a measurement with $80 \%$ cloud coverage. Several events with large signals in pixels of the telescopes were found.}
  \label{clear}
\end{minipage}
\hfill
\end{center}
\end{figure}

\begin{figure}[htbp]
\begin{center}
\includegraphics[width=5.8in]{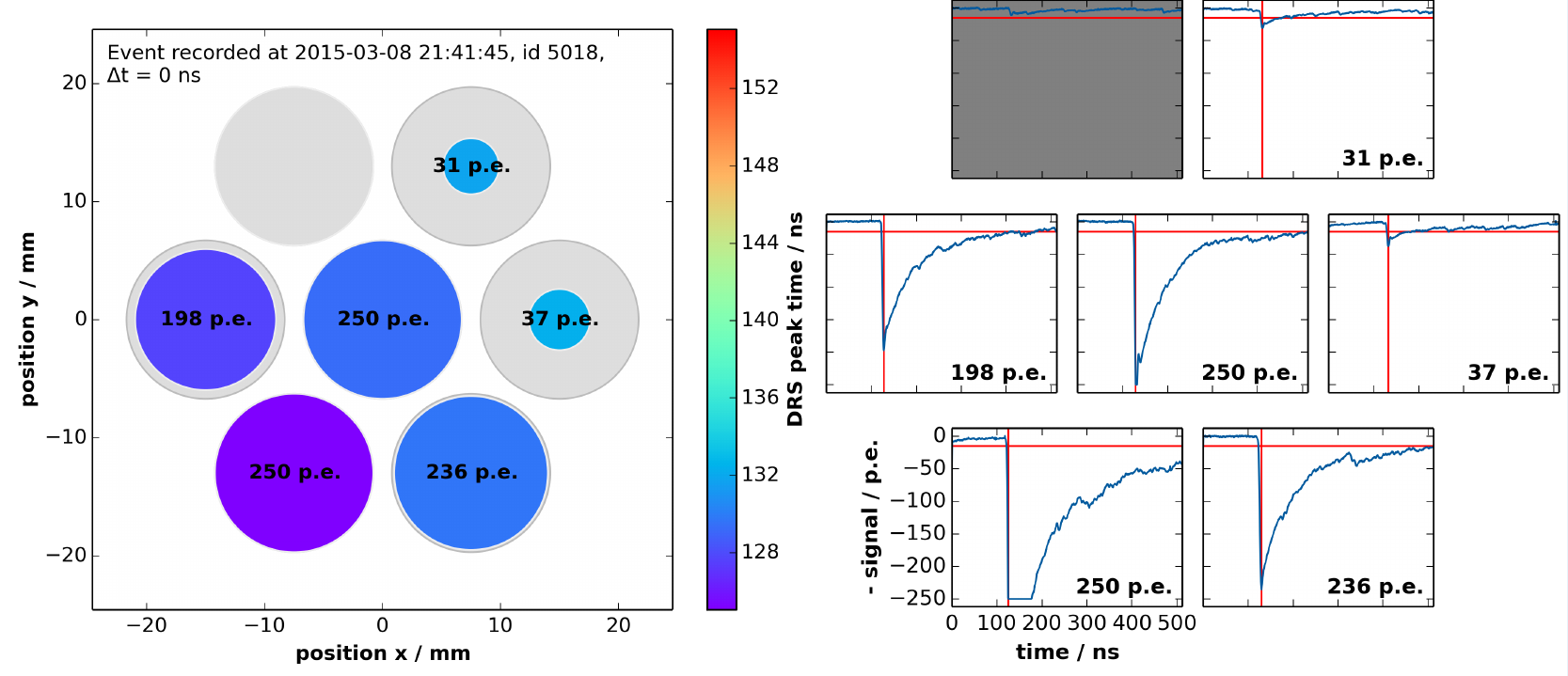}
\caption{Single event measured during field tests in march showing several pulses with sharp rise time and large pulse height in our camera.}
\label{Event1}
\end{center}
\end{figure}

In March 2015, first field tests with the air Cherenkov telescope prototype were made. During two nights, measurements of several thousand seconds have been carried out. The first night was very cloudy. The second was during full moon and relatively clear. Figures \ref{cloudy} and \ref{clear} show the counting rates as a function of largest voltage measured in any of the  pixels of the telescope. Comparing those two, a much larger number of events with a strong signal in one or more pixels have been found in the second night. All the events with high pulses show agreement with the signature as predicted from air shower induce Cherenkov light flashes. Figure \ref{Event1} 
shows an example of these signals. The large size, the sharp rise time, and the close coincidence of the pulses that are only recorded during clear nights are evidences for the detection of air Cherenkov signal.

\section{Summary}

A small robust air Cherenkov telescope has been discussed which is expected to work at the South Pole as an air shower array with the possibility to veto cosmic-ray induced signals in the in-ice detector of IceCube.
Such a telescope with an aperture of $\SI{507}{mm}$ diameter and a field of view of $\ang{12}$ is expected to have an energy threshold of $\SI{100}{TeV}$. With first measurements in the field, the readiness of the system for tests at the South Pole has been proven. 
Detailed tests and simulations of these type of detectors will help to proof the usefulness of air Cherenkov telescopes as detectors at the geographic South Pole.


\begin{thebibliography}{99}
\bibitem{IceCube} A. Achterberg et al., Astropart. Phys., 26 (2006) 155-173, doi: 10.1016/j.astropartphys.2006.06.007.
\bibitem{4YHese} IceCube Coll., \textit{Observation of Astrophysical Neutrinos in Four Years of IceCube Data}, PoS(ICRC2015)1081 these proceedings.
\bibitem{SurV} IceCube Coll., \textit{Motivations and Techniques for a Surface Detector to Veto air showers for Neutrino Astronomy with IceCube in the Southern Sky}, PoS(ICRC2015)1156 these proceedings.
\bibitem{Gen2} IceCube Coll., \textit{IceCube Gen2 High Energy Array}, PoS(ICRC2015)1146 these proceedings.
\bibitem{IceTop} R. Abbasi et al., Nucl. Inst. Meth. A, \textbf{700} (2013) 188-220, doi: 10.1016/j.nima.2012.10.067.
\bibitem{EHE} IceCube Coll., \textit{A search for extremely high energy neutrinos in 6 years of IceCube data}, paper 0463 these proceedings.
\bibitem{PS} IceCube Coll., \textit{Results of neutrino point source searches with 2008-2014 IceCube data above 10 TeV}, PoS(ICRC2015)1047 these proceedings.
\bibitem{SurfaceExtSim} IceCube Coll., \textit{Simulation Studies for a Surface Veto Array to Identify Astrophysical Neutrinos at the South Pole}, PoS(ICRC2015)1070 these proceedings.
\bibitem{Segev} S. Benzvi et. al., \textit{An Estimate of the Live Time of Optical Measurements of Air Showers at the South Pole}, PoS(ICRC2015)568 these proceedings.
\bibitem{Famous} T. Bretz et. al., \textit{{FAMOUS -- A fluorescence telescope using SiPMs}}, PoS(ICRC2015)649 these proceedings.
\bibitem{PowerSupply} J. Schumacher et. al., \textit{Dedicated power supply system for silicon photomultipliers}, PoS(ICRC2015)605 these proceedings.
\bibitem{FACT} A.~Biland et al., JINST \textbf{9} (2014) P10012, http://arxiv.org/abs/1403.5747.
\end{thebibliography}
\end{document}